\documentclass[12pt]{article}
\usepackage{epsfig,amssymb,amsmath,psfrag}

\def \be  {\begin{equation}}
\def \ee  {\end{equation}}
\def \ba  {\begin{eqnarray}}
\def \ea  {\end{eqnarray}}
\def \baa {\begin{eqnarray*}}
\def \eaa {\end{eqnarray*}}
\def \bb  {\begin {thebibliography} }
\def \eb  {\end{thebibliography}}
\def \lab #1 {\label{#1}}

\newcommand\re[1]{(\ref{#1})}

\def \qqqquad {\qquad\qquad}
\def \matrix #1 {\left(\begin{array}{cc} #1 \end{array}\right)}

\def \tr {\mathop{\rm tr}\nolimits}

\def \e  {\mathop{\rm e}\nolimits}
\newcommand\lr[1]{{\left({#1}\right)}}

\newcommand \vev [1] {\langle{#1}\rangle}

\newcommand{\ft}[2]{{\textstyle\frac{#1}{#2}}}
\def \vep {\epsilon}

\def\m{\mu}
\def\C{\Gamma}
\def\del{\partial}
\def\nn{\nonumber}
\def\cM{{\cal M}}
\def\z{\zeta}

\begin{document}

\thispagestyle{empty}
\null\vskip-12pt \hfill  LAPTH-1206/07 \\
\null\vskip-12pt \hfill LPT--Orsay--07--82
\vskip2.2truecm
\begin{center}
\vskip 0.2truecm {\Large\bf
{\Large On planar gluon amplitudes/Wilson loops duality}
}\\
\vskip 1truecm
{\bf J.M. Drummond$^{*}$, J. Henn$^{*}$, G.P. Korchemsky$^{**}$ and E. Sokatchev$^{*}$ \\
}

\vskip 0.4truecm
$^{*}$ {\it Laboratoire d'Annecy-le-Vieux de Physique Th\'{e}orique
LAPTH\footnote{UMR 5108 associ\'{e}e \`{a}
 l'Universit\'{e} de Savoie},\\
B.P. 110,  F-74941 Annecy-le-Vieux, France\\
\vskip .2truecm $^{**}$ {\it
Laboratoire de Physique Th\'eorique%
\footnote{Unit\'e Mixte de Recherche du CNRS (UMR 8627)},
Universit\'e de Paris XI, \\
F-91405 Orsay Cedex, France
                       }
  } \\
\end{center}

\vskip 1truecm 
\centerline{\bf Abstract} \normalsize \noindent

There is growing evidence that on-shell gluon scattering amplitudes in planar
$\mathcal{N}=4$ SYM theory are equivalent to Wilson loops evaluated over contours
consisting of straight, light-like segments defined by the momenta of the
external gluons. This equivalence was first suggested at strong coupling using
the AdS/CFT correspondence and has since been verified  at weak coupling to one
loop in perturbation theory. Here we perform an explicit two-loop calculation of
the Wilson loop dual to the four-gluon scattering amplitude and demonstrate that
the relation holds beyond one loop. We also propose an anomalous conformal Ward
identity which uniquely fixes the form of the finite part (up to an additive
constant) of the Wilson loop dual to four- and five-gluon amplitudes, in complete
agreement with the BDS conjecture for the multi-gluon MHV amplitudes.

\newpage
\setcounter{page}{1}\setcounter{footnote}{0}

\section{Introduction}

Recent studies of scattering amplitudes in $\mathcal{N}=4$ SYM theory have led to a
very interesting all-loop iteration conjecture~\cite{bds05} about the IR finite
part of the color-ordered planar gluon amplitudes. In particular, the four-gluon
amplitude is expected to take the surprisingly simple form
\begin{equation}\label{simpleform}
    \ln {\cal M}_4  = \mbox{[IR divergences]}
    + \frac{\Gamma_{\rm cusp}(a)}{4} \ln^2 \frac{s}{t} + \mbox{const}\ ,
\end{equation}
where $a=g^2N/(8\pi^2)$ is the coupling constant and $s$ and $t$ are the
Mandelstam kinematic variables.\footnote{To avoid the appearance of an imaginary
part in $\ln {\cal M}_4$, it is convenient to choose $s$ and $t$ negative.} The
relation \re{simpleform} suggests that finite corrections to the amplitude ${\cal
M}_4$ exponentiate and the coefficient of  $\ln^2(s/t)$ in the exponent is
determined by the universal cusp anomalous dimension $\Gamma_{\rm
cusp}(a)$~\cite{P80,KR87}. The conjecture \re{simpleform} has been verified up to
three loops in \cite{bds05}. Also, a similar conjecture has been put forward for
maximal helicity violating (MHV) $n$-gluon amplitudes \cite{bds05} and it has
been confirmed for $n=5$ at two loops in \cite{5point}.

Recently, another very interesting proposal has been made \cite{am07} for
studying gluon scattering amplitudes at strong coupling. Within the context of
the AdS/CFT correspondence~\cite{AdS}, and making use of the T-duality
transformation \cite{Kallosh:1998ji}, $\ln {\cal M}_4$ is identified with the
area of the world-sheet of a classical string in AdS space, whose boundary
conditions are determined by the gluon momenta $p^\mu_i$ (with $i=1,\ldots,4$).
As was noticed in \cite{am07}, their calculation is ``mathematically similar'' to
that of the expectation value of a Wilson loop made out of four light-like
segments $(x_i,\ x_{i+1})$ in $\mathcal{N}=4$ SYM at strong
coupling~\cite{M98,Kr02} with dual coordinates $x_i^\mu$ related to the on-shell
gluon momenta by
\be\label{chva}
x_{i,i+1}^\mu \equiv x_i^\mu - x_{i+1}^\mu := p_i^\mu\,.
\ee
Quite remarkably, the resulting stringy expression for $\ln {\cal M}_4$ takes
exactly the same form as in (\ref{simpleform}), with the strong-coupling value of
$\Gamma_{\rm cusp}(a)$ obtained from the semiclassical analysis of
\cite{GPK,KRTT07}. This result constitutes a non-trivial test of the AdS/CFT
correspondence.

In the paper \cite{DKS07} by three of us, we argued that the
duality observed in \cite{am07} between the on-shell gluon amplitudes and
light-like Wilson loops is also valid in planar $\mathcal{N}=4$ SYM at weak
coupling,
\be\label{M=W}
\ln {\cal M}_4 = \ln W\lr{C_4} + O(1/N^2)\,.
\ee
The Wilson loop expectation value on the right-hand side of this relation is
defined as
\begin{equation}\label{W}
    W\lr{C_4} = \frac1{N}\vev{0|{\rm Tr}\, {\rm P} \exp\lr{ig\oint_{C_4} dx^\mu A_\mu(x)}
    |0}\,,
\end{equation}
where $ A_\mu(x)=A_\mu^a(x) t^a$ is a gauge field, $t^a$ are the $SU(N)$
generators in the fundamental representation normalized as $\tr (t^a t^b)=\ft12
\delta^{ab}$ and ${\rm P}$ indicates the ordering of the $SU(N)$ indices along
the integration contour. As in \cite{am07}, the integration contour $C_4$
consists of four light-like segments joining the points $x_i^\mu$ (with
$i=1,2,3,4$) such that $x_i-x_{i+1}=p_i$ coincide with the external on-shell
momenta of the four-gluon scattering amplitude, Eq.~\re{chva}. Going through the
calculation of the light-like Wilson loop \re{W} in the weak coupling limit, we
established the correspondence between the IR singularities of the four-gluon
amplitude and the UV singularities of the Wilson loop and extracted the finite
part of the latter at one loop. Remarkably, our result turned out to be again of
the form (\ref{simpleform}). The relation \re{M=W} admits a natural
generalization to multi-gluon amplitudes. As was later shown in \cite{BHT07}, the
one-loop MHV amplitudes with an arbitrary number $n\ge 5$ of external legs
coincides with the expectation value of the Wilson loop
\be\label{MHV-duality}
\ln {\cal M}_n^{\rm (MHV)} = \ln W\lr{C_n} + O(1/N^2)
\ee
evaluated along a polygonal loop consisting of $n$ light-like segments.

A natural question arises whether the above mentioned correspondence between
gluon scattering amplitudes and light-like Wilson loops holds true to higher
loops. In the present letter, we perform an explicit two-loop calculation of the
light-like Wilson loop $W\lr{C_4}$ entering the right-hand side of \re{M=W} and
match the result into the known expression for the four-gluon amplitude
\cite{Bern:1997nh}. We demonstrate that the two expressions are in a perfect
agreement with each other, thus providing an additional support to the scattering
amplitude/Wilson loop correspondence \re{M=W}.

Finally, in Section \ref{sect3} we discuss the possible consequences of the
conformal symmetry of the Wilson loop on the form of its finite part. Since the
presence of cusps leads to divergences, we should expect that conformal
invariance manifests itself in the form of anomalous Ward identities. We propose
a very simple anomalous conformal-boost Ward identity, which we conjecture to be
valid to all orders. We then show that it uniquely fixes the form of the finite
part (up to an additive constant) of the Wilson loop dual to four- and five-gluon
amplitudes, in complete agreement with the BDS ansatz \cite{bds05} for the
$n$-gluon MHV amplitudes. Furthermore, the $n$-point ansatz of \cite{bds05} also
satisfies this Ward identity for arbitrary $n$. However, starting with six
points, conformal symmetry leaves room for an arbitrary function of the
conformally invariant cross-ratios.

\section{Light-like Wilson loops}

\subsection{One-loop calculation}

In this section, we summarize some general properties of Wilson loops
and review the one-loop calculation of \cite{DKS07}.

The Wilson loop \re{W} is a gauge-invariant functional of the integration contour
$C_4$. We would like to stress that it is defined in configuration space. The
contour $C_4$ consists of four light-cone  segments in Minkowski
space-time between the points $x_i^\mu$ (with $i=1,2,3,4$). It can be
parameterized as
\be\label{C4}
C_4 = \ell_1 \cup \ell_2 \cup \ell_3\cup \ell_4\,,\qquad \ell_i=\{\tau_i x^\mu_i
+ (1-\tau_i) x_{i+1}^\mu|\,0\le \tau_i \le 1\}\,,
\ee
with $x_{i,i+1}^2\equiv (x_i-x_{i+1})^2=0$. Then, to lowest order in the
coupling constant, $W(C_4)$ is given by a sum over double contour integrals
\be\label{double-sum}
 W(C_4) = 1 + \frac12(ig)^2 C_F\sum_{1\le j,\, k\le 4} \int_{\ell_j} dx^\mu \int_{\ell_k} dy^\nu\, G_{\mu\nu}(x-y)  +
 O(g^2)  \,,
\ee
where $C_F=t^a t^a= (N^2-1)/(2N)$ is the quadratic Casimir of $SU(N)$ in the fundamental
representation and $G_{\mu\nu}(x-y)$ is the gluon propagator in the coordinate
representation. In our calculation we employ the Feynman gauge and
introduce dimensional regularization, $D=4-2\epsilon$ (with $\epsilon>0$),
in which case the gluon propagator is given by
\be
G_{\mu\nu}(x)  =  -g_{\mu\nu}\frac{\Gamma(1-\vep)}{4\pi^{2}} (-
x^2+i0)^{-1+\vep}\lr{\mu^2 \pi }^{\vep}\,.
\ee
It proves convenient to redefine the dimensional regularization scale as
\be
\mu^2\pi{\rm e}^\gamma \ \mapsto \ \mu^2
\ee
with the Euler constant $\gamma$ originating from $\Gamma(1-\vep)
=\exp\lr{\gamma\epsilon + O(\epsilon^2)}$.

Let us represent each term in the double sum \re{double-sum} as a Feynman
diagram in which the integration contour is depicted by a double line and the
gluon is attached to the $j$-th and the $k$-th segments (see
Figs.~\ref{all-diags}(a) and
\ref{all-diags}(b)). Then it is easy to see that if both ends of the
gluon are attached to the
same segment, $k=j$, the diagram vanishes due to $G_{\mu\nu}(x-y)x_{j,j+1}^\mu
x_{j,j+1}^\nu \sim x_{j,j+1}^2=0$. If the gluon is attached to two adjacent
segments, $k=j+1$, the corresponding diagram develops a double pole in
$\epsilon$. Finally, for $k=j+2$, the diagram remains finite as
$\epsilon\to 0$.

After a back-of-the-envelope calculation of the one-loop diagrams shown in
Figs.~\ref{all-diags}(a) and \ref{all-diags}(b), we obtain the one-loop
expression for the light-like Wilson loop,
\be\label{1-loop}
\ln W(C_4) = \frac{g^2}{4\pi^2}C_F \left\{ -\frac{1}{\vep^2}
\left[\lr{{-x_{13}^2}{\mu^2}}^\vep+\lr{ {-x_{24}^2}{\mu^2}}^\vep\right] +\frac12
\ln^2\lr{\frac{x_{13}^2}{x_{24}^2}}+\frac{\pi^2}{3}+ O(\vep)\right\}+O(g^4)
\ee
where
\be
x_{jk}^2 \equiv (x_j-x_k)^2
\ee
denotes the distances between the vertices of $C_4$.
The double poles in $\epsilon$ on the right-hand side of \re{1-loop} originate
from the vertex-type Feynman diagram shown in Fig.~\ref{all-diags}(a). They come
from integration over the position of the gluon in the vicinity of the cusp and
have a clear ultraviolet meaning.

Let us substitute \re{1-loop} into the duality relation \re{M=W} and apply
\re{chva} to identify the coordinates $x_{i,i+1}^\mu$ with the on-shell gluon
momenta $p_i^\mu$. This leads to
\be
x_{13}^2 := (p_1+p_2)^2 = s\,,\qquad x_{24}^2 := (p_2+p_3)^2 = t\,,
\ee
where $s$ and $t$ stand for Mandelstam invariants corresponding to the four-gluon
amplitude. Then, we take into account the relation  $C_F=N/2 + O(1/N^2)$ and use
the well-known expression for the one-loop cusp anomalous
dimension~\cite{P80,KR87}, $\Gamma_{\rm cusp}=2 a+ O(a^2)$, to observe that,
firstly, upon identification of the dimensional regularization parameters
\be\label{1-loop-match}
x_{13}^2\,\mu^2 := s/\mu_{\rm IR}^2 \,,\qquad x_{24}^2\,\mu^2 := t/\mu_{\rm
IR}^2\,,\qquad x_{13}^2/x_{24}^2 := s/t
\ee
the UV divergencies of the light-like Wilson loop match the IR divergent part of
the four-gluon scattering amplitude \re{simpleform} and, secondly, the finite
($\sim \ln^2(s/t)$) corrections to these two seemingly different objects coincide
to one loop.

\subsection{Nonabelian exponentiation}

Let us extend the analysis beyond one loop and examine the perturbative expansion
of the light-like Wilson loop \re{W} to order $g^4$. The corresponding
perturbative corrections to $\ln W(C_4)$ come both from subleading terms in the
expansion of the path-ordered exponential \re{W} in powers of the gauge field and
from interaction vertices in the Lagrangian of $\mathcal{N}=4$ SYM theory. As
before, it is convenient to represent $O(g^4)$ corrections as Feynman diagrams.

\begin{figure}[ht]
\psfrag{x1}[cc][cc]{$x_3$} \psfrag{x2}[cc][cc]{$x_2$} \psfrag{x3}[cc][cc]{$x_1$}
\psfrag{x4}[cc][cc]{$x_4$} \psfrag{a}[cc][cc]{(a)} \psfrag{b}[cc][cc]{(b)}
\psfrag{c}[cc][cc]{(c)} \psfrag{d}[cc][cc]{(d)} \psfrag{e}[cc][cc]{(e)}
\psfrag{f}[cc][cc]{(f)} \psfrag{g}[cc][cc]{(g)} \psfrag{h}[cc][cc]{(h)}
\psfrag{i}[cc][cc]{(i)} \psfrag{j}[cc][cc]{(j)} \psfrag{k}[cc][cc]{(k)}
\psfrag{l}[cc][cc]{(l)} \centerline{{\epsfysize14cm \epsfbox{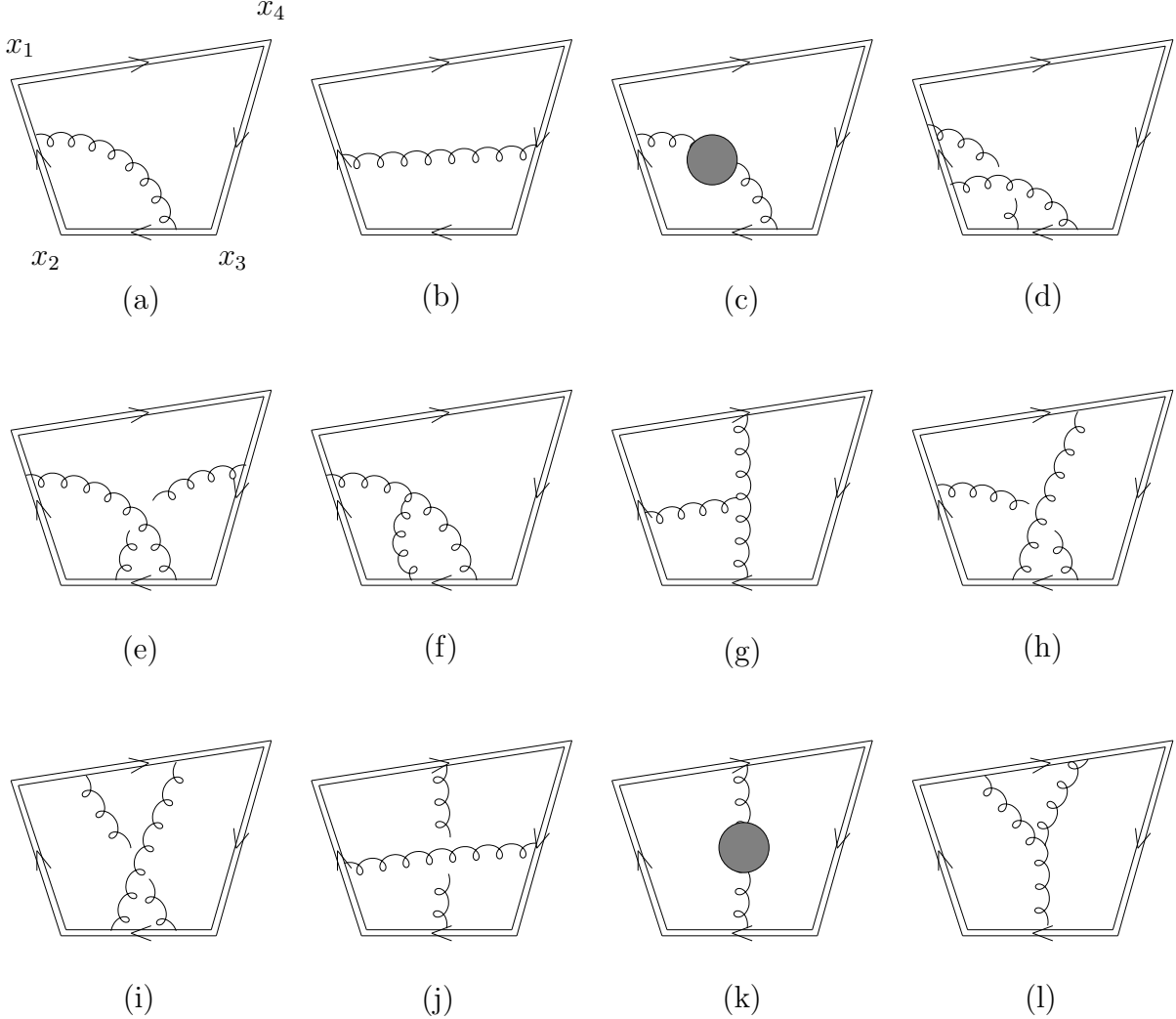}}}
\caption[]{\small The Feynman diagrams contributing to $\ln {W(C_4)}$ to two
loops. The double line depicts the integration contour $C_4$, the wiggly line the
gluon propagator and the blob the one-loop polarization operator.}
\label{all-diags}
\end{figure}

Let us classify all possible $O(g^4)$ diagrams according to the order in the
expansion of the path-ordered exponential in \re{W} in the gauge field, or
equivalently, to the number of gluons $n_g=2,3,4$ attached to the integration
contour (depicted by a double line in Fig.~\ref{all-diags}). It is easy to see
that for $n_g=4$ the relevant diagrams cannot contain interaction vertices (see
Figs.~\ref{all-diags}(d), \ref{all-diags}(e), \ref{all-diags}(h),
\ref{all-diags}(i) and \ref{all-diags}(j)). Moreover, if one of the gluons is
attached by both legs to the same light-like segment, the diagram vanishes in the
Feynman gauge for the same reason as at one loop. For $n_g=3$, three gluons
attached to the integration contour can be only joined together
through a
three-gluon vertex $V_{\mu_1\mu_2\mu_3}$ (see Figs.~\ref{all-diags}(f),
\ref{all-diags}(g) and \ref{all-diags}(l)). In addition, if all three gluons are
attached to the same light-like segment, the diagram vanishes by virtue of
$V_{\mu_1\mu_2\mu_3}x_{j,j+1}^{\mu_1} x_{j,j+1}^{\mu_2}x_{j,j+1}^{\mu_3} \sim
x_{j,j+1}^2 =0$. Finally, for $n_g=2$, the corresponding diagrams take the same
form as the one-loop diagrams with the only difference being that the gluon propagator
gets ``dressed'' by $O(g^2)$ corrections (see Figs.~\ref{all-diags}(c) and
\ref{all-diags}(k)). The latter corrections come both from tadpole diagrams which
vanish in dimensional regularization and from gauge
fields/gauginos/scalars/ghosts propagating inside the loop.

Another dramatic simplification occurs after we take into account the nonabelian
exponentiation property of Wilson loops~\cite{Gatheral}. In application to the
light-like Wilson loop \re{W} in $\mathcal{N}=4$ SYM theory, it can be formulated
as follows,
\be\label{exponentiation}
W(C_4) = 1+ \sum_{n=1}^\infty \lr{\frac{g^2}{4\pi^2}}^n W^{(n)} =
\exp\left[{\sum_{n=1}^\infty \lr{\frac{g^2}{4\pi^2}}^n c^{(n)} w^{(n)}}\right]\,.
\ee
Here $W^{(n)}$ denote perturbative corrections to the Wilson loop, while
$c^{(n)}w^{(n)}$ are given by the contribution to $W^{(n)}$ with the ``maximally
nonabelian'' color factor $c^{(n)}$. To the first few orders in $n=1,2,3$ the
maximally nonabelian color factor takes the form $c^{(n)}=C_F N^{n-1}$ but
starting from $n=4$ loops it is not expressible in terms of simple Casimir
operators. We deduce from \re{exponentiation} that
\be\label{exp-example}
W^{(1)}=C_F w^{(1)}\,,\qquad W^{(2)} = C_F N w^{(2)}+ \frac12C_F^2
\lr{w^{(1)}}^2\,, \qquad \ldots
\ee
Then it follows from \re{exp-example} that the correction to $W^{(2)}$
proportional to $C_F^2$ is uniquely determined by the one-loop correction
$W^{(1)}$. This property allows us to reduce significantly the number of relevant
two-loop diagrams.

Let us examine the color factors corresponding to the various two-loop Feynman
diagrams. Following the classification of diagrams presented at the beginning of
this section, we find that nonvanishing diagrams with $n_g=2$ and $n_g=3$ are
accompanied by the same color factor $C_F N$ and, therefore, contribute to
$w^{(2)}$. For $n_g=4$, that is for diagrams without interaction vertices
(abelian-like diagrams), the color factor equals $t^a t^a t^b t^b=C_F^2$ and $t^a
t^b t^a t^b=C_F(C_F-N/2)$ for planar and nonplanar diagrams, respectively.
Applying nonabelian exponentiation \re{exp-example} we obtain that the planar
$n_g=4$ diagrams do not contribute to $w^{(2)}$ and, therefore, they can be
safely neglected. At the same time, to define the contribution of nonplanar
$n_g=4$ diagrams to $w^{(2)}$, we have to retain the maximally nonabelian
contribution only. This can be easily done by replacing the color factor of the
diagram by
$C_F(C_F-N/2)\to -C_F N/2$. %
\footnote{Notice that $C_F(C_F-N/2)$ is subleading in the multi-color limit. It
is amusing that the planar expression for the Wilson loop $W(C_4)$ is given by an
exponential which receives corrections from nonplanar diagrams.}

To summarize, we list in Fig.~\ref{all-diags} all nonvanishing, two-loop diagrams
of different topology contributing to $w^{(2)}$.

\subsection{Two-loop calculation}

In order to compute the two-loop Feynman diagrams shown in Fig.~\ref{all-diags}
we employ the technique described in detail in Refs.~\cite{K88,KK92}.
Furthermore, the rectangular light-like Wilson loop under consideration has been
already calculated to two loops in \cite{KK92} in the so-called forward limit
$x_{12}^\mu=-x_{34}^\mu$, or equivalently $x_{13}^2=-x_{24}^2$, in which the
contour $C_4$ takes the form of a rhombus. The vertex-like diagrams shown in
Fig.~\ref{all-diags}(c), \ref{all-diags}(d) and \ref{all-diags}(f) only depend on
the distances $x_{13}^2$ and we can take the results from \cite{KK92}. The
Feynman diagrams shown in Fig.~\ref{all-diags}(h) -- \ref{all-diags}(l) are
proportional to scalar products $(x_{12}\cdot x_{34})$ and/or $(x_{23}\cdot
x_{14})$ and, therefore, they vanish in the limit $x_{12}^\mu=-x_{34}^\mu$ due to
$x_{i,i+1}^2=0$. For generic forms of $C_4$, these diagrams have to be calculated
anew. For the same reason, we have also to reexamine the contribution of diagrams
shown in Figs.~\ref{all-diags}(e) and \ref{all-diags}(g). Finally, the
calculation in Ref.~\cite{KK92} was performed within conventional dimensional
regularization (DREG) scheme. To preserve supersymmetry, we have to use instead
the dimensional reduction (DRED) scheme. The change of scheme only affects the
diagrams shown in Figs.~\ref{all-diags}(c) and \ref{all-diags}(k) which involve
an internal gluon loop~\cite{BGK03}.

The results of our calculation can be summarized as follows. Thanks to nonabelian
exponentiation, the two-loop expression for the (unrenormalized) light-like Wilson
loop can be represented as
\be\label{W-decomposition}
\ln W(C_4) = \frac{g^2}{4\pi^2}C_F w^{(1)}  +  \lr{\frac{g^2}{4\pi^2}}^2 C_F N
w^{(2)}  + O(g^6)\,.
\ee
According to \re{1-loop} the one-loop correction $w^{(1)}$ is given by
\be\label{w1}
w^{(1)}=-\frac{1}{\vep^2} \left[\lr{{-x_{13}^2}\,{\mu^2}}^\vep+\lr{
{-x_{24}^2}\,{\mu^2}}^\vep\right] +\frac12
\ln^2\lr{\frac{x_{13}^2}{x_{24}^2}}+\frac{\pi^2}{3}+ O(\epsilon)\,.
\ee
The two-loop correction $w^{(2)}$ is given by a sum over the individual diagrams
shown in Fig.~\ref{all-diags} plus crossing symmetric diagrams. It is convenient
to expand their contributions in powers of $1/\epsilon$ and separate the UV divergent
and finite parts as follows
\be\label{para}
w^{(2)} = \sum_{\alpha}  \left[(-x_{13}^2\, \mu^2)^{2\epsilon}+ (-x_{24}^2\,
\mu^2)^{2\epsilon}\right] \bigg\{\frac1{\epsilon^4}A_{-4}^{(\alpha)} +
\frac1{\epsilon^3}A_{-3}^{(\alpha)} + \frac1{\epsilon^2}A_{-2}^{(\alpha)} +
\frac1{\epsilon}A_{-1}^{(\alpha)}\bigg\} + A_0^{(\alpha)} + O(\epsilon)\,,
\ee
where the sum goes over the two-loop Feynman diagrams shown in
Fig.~\ref{all-diags}(c)--(l). Here $A^{(\alpha)}_{-n}$ (with $0\le n \le 4$) are
dimensionless functions of the ratio of distances $x_{13}^2/x_{24}^2$. Making use
of \re{para}, we can parameterize the contribution of each individual diagram to
the Wilson loop by the set of coefficient functions $A^{(\alpha)}_{-n}$:

\begin{itemize}

\item UV divergent $O(1/\epsilon^4)$ terms only come from the two Feynman diagrams
shown in Figs.~\ref{all-diags}(d) and \ref{all-diags}(f)
\be\label{A4}
A_{-4}^{\rm (d)}=-\frac1{16}\,,\qqqquad A_{-4}^{\rm (f)}= \frac1{16}
\ee

\item  UV divergent $O(1/\epsilon^3)$ terms only come from the two Feynman diagrams
shown in Figs.~\ref{all-diags}(c) and \ref{all-diags}(f)
\be\label{A3}
A_{-3}^{\rm (c)}=\frac1{8}\,,\qqqquad A_{-3}^{\rm (f)}=-\frac1{8}
\ee

\item   UV divergent $O(1/\epsilon^2)$ terms only come from the Feynman diagrams
shown in Figs.~\ref{all-diags}(c)-- \ref{all-diags}(g)
\be\label{A2}
A_{{-2}}^{\rm (c)}=\frac14\,,\qquad A_{{-2}}^{\rm (d)}=-{\frac
{\pi^2}{96}}\,,\qquad A_{{-2}}^{\rm (e)}=-\frac{\pi^2}{24} \,,\qquad
A_{{-2}}^{\rm (f)}=-\frac14+{\frac {5}{96}}\,{\pi }^{2}\,,\qquad A_{{-2}}^{\rm
(g)}= \frac{{\pi }^{2}}{48}
\ee

\item  UV divergent $O(1/\epsilon^1)$ terms come from the Feynman diagrams shown in
Figs.~\ref{all-diags}(c)--\ref{all-diags}(h),\ref{all-diags}(k) and
\ref{all-diags}(l)
\begin{align} \nonumber
&  A_{{-1}}^{\rm (c)}=\frac12+\frac{{\pi }^{2}}{48}\,,  & & A_{{-1}}^{\rm
(d)}=-\frac1{24}\zeta_3\,, &
\\ \nonumber
 & A_{{-1}}^{\rm (e)}=\frac12 \zeta_3\,, & &
A_{{-1}}^{\rm (f)}=-\frac12-\frac{{\pi }^{2}}{48} +{\frac {7}{24}} \zeta_3\,, &
\\ \nonumber
& A_{{-1}}^{\rm (g)}=-\frac18 M_{{2}}+\frac18 \zeta_3\,, & & A_{{-1}}^{\rm
(h)}=\frac14 M_{{2}}\,,  &
\\ \label{A1}
 & A_{{ -1}}^{\rm (k)}=\frac14 M_{{1}}\,,
& & A_{{-1}}^{\rm (l)}=-\frac14 M_{{1}}-\frac18 M_{{2}} &
\end{align}

\item  Finite $O(\epsilon^0)$ terms come from all Feynman diagrams shown in
Figs.~\ref{all-diags}(c)--\ref{all-diags}(l)
\begin{align} \nonumber
&  A_{{0}}^{\rm (c)}= 2+\frac{{\pi }^{2}}{12}+\frac16\zeta_3\,, & & A_{{0}}^{\rm
(d)}=-{\frac {7}{2880}}\,{\pi }^{4} \,, &
\\ \nonumber
& A_{{0}}^{\rm (e)}=-\frac{{\pi }^{2}}{12} M_{{1}}-{ \frac {49}{720}}\,{\pi }^{4}
\,, & & A_{{0}}^{\rm (f)}=-2-\frac{{\pi }^{2}}{12}+{\frac {119 }{2880}}\,{\pi
}^{4}-\frac16 \zeta_3 \,, &
\\ \nonumber
& A_{{0}}^{\rm (g)}=\frac1{24} M_ {{1}}^{2}-\frac14 M_{{3}}+{\frac
{7}{360}}\,{\pi }^{4} \,, & & A_{{0}}^{\rm (h)}= \frac18 M _{{1}}^{2}+\frac38
M_{{3}}+\frac{\pi^2}8 M_{{1}} \,,  &
\\ \nonumber
& A_{{0}}^{\rm (i)}=-\frac1{24} M_{{1 }}^{2} \,, &&  A_{{0}}^{\rm
(j)}=-\frac1{8}M_{{1}}^{2} &
\\ \label{A0}
& A_{{0}}^{\rm (k)}= M_{{1}}+\frac12\,M_{{2}}\,, & & A_{{0}}^{\rm (l)}=
-M_{{1}}+\frac{\pi^2}{24} M_{{1}}-\frac12 M_{{2}}-\frac18\,M_{{3}}\,.  &
\end{align}

\end{itemize}
Here the notation was introduced for the integrals $M_i=M_i(x_{13}^2/x_{24}^2)$
\ba \nonumber
M_{1} &=& \int_{0}^{1}\frac{d\beta}{\beta-\bar{\alpha}}
\ln\left(\frac{\bar{\alpha}
\bar{\beta}}{\alpha \beta}\right) = -\frac{1}{2} \left[ \pi^2 +
\ln^2\left(\frac{x_{13}^2}{x_{24}^2}\right) \right],
\\ \nonumber
M_{2} &=& \int_{0}^{1}\frac{d\beta}{\beta-\bar{\alpha}}
\ln\left(\frac{\bar{\alpha}
\bar{\beta}}{\alpha \beta}\right) \ln(\beta
\bar{\beta})\,,
\\ \label{M-integrals}
M_{3} &=& \int_{0}^{1}\frac{d\beta}{\beta-\bar{\alpha}}
\ln\left(\frac{\bar{\alpha}
\bar{\beta}}{\alpha \beta}\right) \ln^2(\beta
\bar{\beta})\,,
\ea
with $\bar \beta=1-\beta$, $\bar\alpha=1-\alpha$ and
$\bar\alpha/\alpha=x_{13}^2/x_{24}^2$. We do not need the explicit expressions
for the integrals $M_{2}$ and $M_{3}$ since, as we will see shortly, the
contributions proportional to $M_{2}$ and $M_{3}$ cancel in the sum of all
diagrams (for completeness, they can be found in the Appendix). In what follows
it will be only important that the integrals \re{M-integrals} vanish in the
forward limit $x_{13}^2=-x_{24}^2$.

We would like to stress that the above results were obtained in Feynman
gauge. Despite the fact that the contribution of each individual
Feynman diagram
to the light-like Wilson loop (or equivalently, the
$A_{-n}^{(\alpha)}-$functions) is gauge-dependent, their sum is gauge-invariant.
Then, we substitute the obtained expressions for the coefficient functions,
Eqs.~\re{A4} -- \re{A0}, into \re{para} and finally arrive at the following remarkably
simple expression for the two-loop correction,
\be\label{para1}
w^{(2)} =  \left[(-x_{13}^2\,\mu^2)^{2\epsilon}+
(-x_{24}^2\,\mu^2)^{2\epsilon}\right] \bigg\{ \epsilon^{-2} \frac{\pi^2}{48} +
\epsilon^{-1}\frac{7}{8}\zeta_3\bigg\} -
\frac{\pi^2}{24}\ln^2\lr{\frac{x_{13}^2}{x_{24}^2}} - \frac{37}{720}\pi^4 +
O(\epsilon)\,.
\ee
This relation is one of the main results of the paper. We verify that in the
forward limit, for ${x_{13}^2}=-{x_{24}^2}$, this relation is in agreement with
the previous calculations of Refs.~\cite{KK92}.

The following comments are in order. Arriving at \re{para1} we notice that the
leading UV divergent $O(1/\epsilon^4)$ and $O(1/\epsilon^3)$  terms  cancel in
the sum of all diagrams. According to \re{A2}, the coefficients in front of
$1/\epsilon^2$ are given by a sum of a rational number and $\pi^2-$term. The
rational terms cancel in the sum of all diagrams. As a consequence, the residue
of the double pole in $\epsilon$ of $w^{(2)}$ in Eq.~\re{para1} is proportional
to $\zeta_2$. In a similar manner, the residue of $w^{(2)}$ at the single pole in
$\epsilon$ is proportional to $\zeta_3$ and this comes about as the result of a
cancelation between various terms in \re{A1} containing rational numbers,
$\pi^2-$terms as well as the integrals $M_1$ and $M_2$. The most striking
simplifications occur in the sum of finite $O(\epsilon^0)$ terms \re{A0}. We find
that the integrals $M_2$, $M_3$, $M_1^2$ as well as the rational corrections and
the terms proportional to $\pi^2$ and $\zeta_3$ cancel in the sum of all diagrams
leading to $-\ft7{720}\pi^4+\ft1{12}\pi^2 M_1$.

We would like to stress that the two-loop expression \re{para1} verifies the
``maximal transcendentality principle'' in $\mathcal{N}=4$ SYM~\cite{KLOV}. Let
us assign transcendentality $1$ to a single pole $1/\epsilon$. Then, it is easy
to see from \re{para1} that the coefficient in front of $1/\epsilon^n$ (including
the finite $O(\epsilon^0)$ term!) has transcendentality equal to $4-n$. In this
way, each term in the two-loop expression \re{para1} has transcendentality $4$.
For the same reason, the one-loop correction to the Wilson loop, Eq.~\re{w1} has
transcendentality $2$. Generalizing this remarkable property to higher loops
in planar $\mathcal{N}=4$ SYM, we expect that the perturbative
correction to the
Wilson loop \re{W} to order $O(g^{2n})$ should have transcendentality
$2n$.

Notice that in our calculation of the two-loop Wilson loop we did not
rely on the
multi-color limit. In fact, due to the special form of the maximally
nonabelian color
factors, $c^{(n)} = C_F N^{n-1}$ the relation \re{W-decomposition} is exact for
arbitrary $N$. As was already mentioned, these color factors become more
complicated starting from $n=4$ loops, where we should expect terms subleading in
$N$.

\section{Duality relation to two loops}\label{sect3}

We combine the one-loop and two-loop corrections to the Wilson loop,
Eqs.~\re{w1} and \re{para1}, respectively, and rewrite the relation
\re{W-decomposition} in the multi-color limit by splitting it into UV divergent
and finite parts as
\be\label{W-final}
\ln W(C_4) = D_4(-x_{13}^2\,\mu^2) + D_4(-x_{24}^2\,\mu^2) +
\mathcal{F}_4(x_{13}^2/x_{24}^2)\,.
\ee
The divergent part is given by the sum over poles
\be\label{D-part}
D_4(-x_{13}^2\,\mu^2) = -\frac{a}{\epsilon^2}
(-x_{13}^2\,\mu^2)^{\epsilon}+a^2
  (-x_{13}^2\,\mu^2)^{2\epsilon} \bigg\{
\frac1{\epsilon^{2}} \frac{\pi^2}{24} + \frac1{\epsilon}\frac{7}{4}\zeta_3\bigg\}
+ O(a^3)
\ee
and similarly for $D_4(-x_{24}^2\,\mu^2)$, while the finite part is given
by
\be\label{F-part}
\mathcal{F}_4(x_{13}^2/x_{24}^2) = \frac{1}4 \ln^2
\left(\frac{x_{13}^2}{x_{24}^2}\right) \left[2a-\frac{\pi^2}3\,a^2+O(a^3) \right]
+ \lr{ \frac{\pi^2}{3}a-\frac{37\pi^4}{360}a^2 + O(a^3)}
\ee
with the coupling constant $a=g^2N/(8\pi^2)$.

We remind the reader that the poles in $\epsilon$ in the expressions for
$D_4(-x_{13}^2\,\mu^2)$ and $D_4(-x_{24}^2\,\mu^2)$ have an UV origin and they appear
due to the fact that the integration contour $C_4$ in the definition of the
Wilson loop \re{W} has light-like cusps. This implies that the dependence of
$W(C_4)$ on the UV cut-off $\mu^2$ should be described by a renormalization
group equation. For the light-like Wilson loop {under consideration}, such an
equation was derived in Refs.~\cite{KK92}:
\be\label{Ward}
\frac{\partial}{\partial\ln\mu^2} \ln W(C_4)= -\frac12 \Gamma_{\rm cusp}
(a)\ln\lr{ x_{13}^2 x_{24}^2 {\mu^4}} - \Gamma(a) -
\frac1{\vep}\int_0^a\frac{da'}{a'}\Gamma_{\rm cusp} (a') + O(\vep)\,,
\ee
where $\Gamma_{\rm cusp}(a)$ is the cusp anomalous dimension and $\Gamma(a)$ is
the so-called collinear anomalous dimension. As a nontrivial check of our
calculation, we verify that the obtained two-loop expression for $\ln W(C_4)$,
Eq.~\re{W-final}, satisfies the evolution equation \re{Ward}. Furthermore, we
find the following two-loop expressions for the anomalous dimensions:
\be\label{cusp-2loop}
\Gamma_{\rm cusp}(a) = 2a -\frac{\pi^2}3 a^2 + O(a^3)\,, \qquad \Gamma(a) = -
7\zeta_3 a^2 + O(a^3)\,.
\ee
These relations are in agreement with the known results~\cite{KR87,BGK03,DKS07}.
Given the fact that $\ln W(C_4)$ obeys the maximal transcendentality principle,
it is not surprising that the two anomalous dimensions have the same property.

We are now in a position to test the duality relation \re{M=W}. According to
\re{M=W}, the two-loop expression for the light-like Wilson loop \re{W-final}
should match the two-loop expression for the four-gluon scattering amplitude
\re{simpleform} upon appropriate identification of the kinematic variables and
dimensional regularization parameter. As was explained in \cite{DKS07}, this can
be done in two steps. Firstly, we identify the dual coordinates
$x_i^\mu$ with the
on-shell gluon momenta according to the rule \re{chva}. Secondly, we
replace the
dimensionless combinations $x_{jk}^2 \mu^2$ of the Wilson loop by the
following kinematic invariants of the gluon amplitude:
\be\label{2-loop-match}
x_{13}^2\,\mu^2 := s/\mu_{\rm IR}^2 \e^{\gamma(a)} \,,\qquad x_{24}^2\,\mu^2 :=
t/\mu_{\rm IR}^2\e^{\gamma(a)} \,,\qquad x_{13}^2/x_{24}^2 : = s/t\,.
\ee
Here $\mu_{\rm IR}^2$ is the parameter of dimensional regularization
which plays the r\^ole of an IR cut-off in the four-gluon amplitude
while $s$ and $t$ are the
Mandelstam variables. Compared to the corresponding one-loop relation
\re{1-loop-match}, Eq.~\re{2-loop-match} involves an additional
function of the coupling constant $\gamma(a)$. Its expression to one
loop can be found in \cite{DKS07}.

Let us start with the divergent part of the Wilson loop defined in \re{D-part}
and apply the relations \re{2-loop-match}. Making use of the evolution equation
\re{Ward} and following the analysis of \cite{DKS07}, it is straightforward to
verify that, for arbitrary values of the kinematic invariants $s$ and $t$, the UV
divergent part of the Wilson loop coincides with the IR divergent part of the
four-gluon scattering amplitude. Finally, we compare the finite part of the
Wilson loop, Eq.~\re{F-part}, with the finite part of the four-gluon amplitude,
Eq.~\re{simpleform}. Substituting the cusp anomalous dimension in \re{simpleform}
by its two-loop expression \re{cusp-2loop}, we find that the finite $\sim
\ln^2(s/t)$ parts in the two relations coincide indeed! This constitutes yet
another confirmation that the duality between gluon amplitudes and Wilson loops
holds true in planar $\mathcal{N}=4$ SYM both at weak and strong coupling.

In our analysis, we restricted ourselves to four-gluon scattering amplitudes. It
is straightforward to extend these considerations to multi-gluon MHV amplitudes.
According to \re{MHV-duality}, the duality relation in that case involves a
light-like Wilson loop evaluated along an $n$-sided polygon, where $n$ matches
the number of external gluons. To one-loop level, the relation has been
established in \cite{BHT07}. To two loops, the corresponding Wilson loop can be
calculated by the method described above. It would be interesting to check
whether the correspondence works in this case as well.

\section{A conformal Ward identity for the Wilson loop}

In this section, we examine the possible consequences of the conformal invariance
of ${\cal N}=4$ SYM on the functional form of the Wilson loop $W(C_n)$ entering
the duality relations \re{M=W} and \re{MHV-duality}. First, we remark that a
straight light-like segment $\ell_j$ of the type \re{C4} maps to another straight
light-like segment under the $SO(2,4)$ conformal transformations (most easily
seen by performing conformal inversion). This means that the entire $n$-sided
polygonal Wilson loop $C_n$ with light-like edges maps into a similar contour
under such transformations. Furthermore, the gauge field $A_\m(x)$ transforms
with conformal weight one and hence the Wilson loop $W(C_n)$ is invariant but for
the change of the contour.

If we were dealing with a finite object, we would then apply the usual procedure
of deriving Ward identities (exploiting the conformal invariance of the classical
action) to deduce that the dilatation and special conformal generators annihilate
the Wilson loop expectation value (see e.g. \cite{Braun:2003rp} and references
therein). However, the Wilson loop $W(C_n)$ is in fact divergent because of the
presence of cusps on the contour. Therefore we need to introduce a regulator,
e.g. dimensional, which breaks conformal invariance. As a consequence, we have to
expect that the conformal Ward identity will receive an anomalous
contribution~\cite{Mueller:1993hg}. We have investigated in detail how this works
at the one-loop perturbative level. The resulting anomalous conformal Ward
identity has a very simple and suggestive structure, which allows us to make a
conjecture about its all-loop form. Leaving the detailed study and proof of this
identity to a further publication, here we examine its consequences. Quite
remarkably, the proposed Ward identity fixes uniquely the four- and five-point
Wilson loops (up to an additive constant), in perfect agreement with the
all-order ansatz of \cite{bds05} and the conjectured equivalence with MHV
amplitudes. We also show that the conjectured form of the $n$-point amplitude of
\cite{bds05} does indeed verify the identity for any $n$.

In a close analogy with \re{W-final}, the Wilson loop $W_n \equiv W(C_n)$ can be
split into divergent and finite parts
\be\label{cf.}
\ln  W_n  = \ln Z_n + \frac{1}{2} \C_{\rm cusp}(a) F_n + O(\epsilon),
\ee
where the cusp anomalous dimension is put in front of $F_n$ for later
convenience. Here $Z_n$ is a UV divergent factor dependent on the dimensional
regularisation scale $\mu$, the regulator $\epsilon$ as well as the points $x_i$
describing the contour $C_n$. It takes the following form~\cite{KK92}
\begin{equation}\label{5}
\ln Z_n = -\frac{1}{4}\sum_{l=1}^\infty a^l\left(\frac{\Gamma_{\rm
cusp}^{(l)}}{(l\epsilon)^2} + \frac{\Gamma^{(l)}}{l\epsilon} \right)\sum_{i=1}^n
(-x_{i,i+2}^2\,\mu^{2})^{l\epsilon},
\end{equation}
where $\C_{\rm cusp}(a) = \sum_{l=1}^{\infty} a^l \C^{(l)}_{\rm cusp}$ and $\C(a)
= \sum_{l=1}^{\infty} a^l \C^{(l)}$. The term $F_n$ is the finite contribution
dependent only on the points $x_i$, as we argue below.

What can we say about the finite part $F_n$ on general grounds? First of all, it
is manifestly rotation and translation invariant, i.e. it is a function of the
variables  $x_{ij}^2$. Furthermore, due to the special choice of the divergent
part $Z_n$ (\ref{5}), where the scale $\mu$ appears in combination with the
coupling,  $a\mu^{2\epsilon}$, we expect that $F_n$ at $\epsilon=0$ should be
independent of $\mu$. Then it is easy to show that the dilatation Ward identity
for the Wilson loop does not produce an anomalous dimension term for $F_n$ and it
is reduced to the trivial form
\be
D\, F_n \equiv \sum_{i=1}^n \lr{x_i \cdot \partial_{x_i}} F_n =0\ .
\ee
This just has the consequence that $F_n$ is a function of the dilatation
invariant ratios of the $x_{ij}^2$. However, an anomaly does appear in the
special conformal Ward identity for $F_n$. As mentioned earlier, based on our
experience with one loop, we propose the following general form,
\be
K^\mu \, F_n = \sum_{i=1}^n(x_{i}^\mu + x_{i+2}^\mu -2 x_{i+1}^\mu)\,  \ln
x_{i,i+2}^2 = \sum_{i=1}^n x_{i,i+1}^\mu \ln \Bigl(
\frac{x_{i,i+2}^2}{x_{i-1,i+1}^2} \Bigr) \ , \label{cwi}
\ee
where the special conformal generators (``boosts") are given by the standard
expression
\be\label{K}
K^\m = \sum_{i=1}^n  2 x_i^\m (x_i \cdot \del_{x_i}) - x_i^2 \del_{x_i}^\m
\ee
and a periodicity condition $x_i = x_{i+n}$ is assumed.

Let us now examine the consequences of the conformal Ward identity \re{cwi} for
the finite part of the Wilson loop $W_n$. We find that the cases of $n=4$ and
$n=5$ are special because here the Ward identity (\ref{cwi}) has a unique
solution up to an additive constant. The solutions are, respectively,
\begin{align}
F_4 &= \frac{1}{2} \ln^2\Bigl(\frac{x_{13}^2}{x_{24}^2}\Bigr) + \text{
 const } \,, \nn\\
F_5 &= - \frac{1}{4} \sum_{i=1}^5 \ln
 \Bigl(\frac{x_{i,i+2}^2}{x_{i,i+3}^2}\Bigr) \ln
 \Bigl(\frac{x_{i+1,i+3}^2}{x_{i+2,i+4}^2}\Bigr) + \text{ const }\ .
\label{remarkable}
\end{align}
This is easy to check by making use of the action of the special conformal
generators \re{K} on the logarithmic terms,
\be
K^\m \ln \Bigl(\frac{x_{i,j}^2}{x_{k,l}^2}\Bigr) = 2(x_{i}^\m + x_j^\m -
 x_k^\m - x_l^\m)\ .
\label{Klog}
\ee
We find that, upon identification of the kinematic invariants
\be\label{xx}
x_{k,k+r}^2 := (p_k+\ldots  + p_{k+r-1})^2\,,
\ee
the relations (\ref{remarkable}) are exactly the functional forms of the ansatz
of \cite{bds05} for the finite parts of the four- and five-point MHV amplitudes.

The reason that the functional form of $F_4$ and $F_5$ is fixed up to an additive
constant is that there are no conformal invariants one can build from four or
five points $x_i$ with light-like separations $x_{i,i+1}^2=0$. Starting from six
points there are conformal invariants in the form of cross-ratios,
\be\label{crr}
K^\m \lr{\frac{x_{i,j}^2 x_{k,l}^2}{x_{i,l}^2 x_{j,k}^2}} = 0\,.
\ee
For example, at six points there are three of them,
\be
u_1 = \frac{x_{13}^2 x_{46}^2}{x_{14}^2 x_{36}^2}, \qquad u_2 = \frac{x_{24}^2
x_{15}^2}{x_{25}^2 x_{14}^2}, \qquad u_3 = \frac{x_{35}^2 x_{26}^2}{x_{36}^2
x_{25}^2}\ .
\ee
Hence the general solution of the Ward identity at six points and higher will
contain an arbitrary function of the conformal cross-ratios.

We now wish to show that the ansatz of \cite{bds05} for the finite part of the
$n$-point MHV gluon amplitudes in ${\cal N}=4$ SYM does satisfy our proposed
conformal Ward identity (\ref{cwi}).  The ansatz of \cite{bds05} for the
logarithm of the ratio of the amplitude to the tree amplitude reads (cf.
(\ref{cf.}))
\be\label{ansatz1}
\ln \cM_n^{\rm (MHV)} = \ln Z_n + \frac{1}{2} \C_{\rm cusp}(a) F_n + C_n
+O(\epsilon)\,,
\ee
where $Z_n$ is the IR divergent part, $F_n$ is the finite part depending on the
Mandelstam variables and $C_n$ is the constant term. At four points the proposed
form of the finite part is  (using the notation \re{xx})
\be
F_4 = \frac{1}{2} \ln^2\Bigl(\frac{x_{13}^2}{x_{24}^2}\Bigr) + 4 \z_2\ ,
\label{F4}
\ee
while for $n\geq 5$ it is
\be
F_n = \frac{1}{2} \sum_{i=1}^n g_{n,i}\,,\quad
g_{n,i} = - \sum_{r=2}^{\lfloor \tfrac{n}{2} \rfloor -1} \ln
\Bigl(\frac{x_{i,i+r}^2}{x_{i,i+r+1}^2}\Bigr) \ln
\Bigl(\frac{x_{i+1,i+r+1}^2}{x_{i,i+r+1}^2}\Bigr) + D_{n,i} + L_{n,i} +
\frac{3}{2} \z_2\ .
\ee
The functions $D_{n,i}$ and $L_{n,i}$ depend on whether $n$ is odd or even. For
$n$ odd, $n=2m+1$ they are
\begin{align} \label{di1}
D_{n,i} &= - \sum_{r=2}^{m-1} {\rm Li}_2 \Bigl(1 - \frac{x_{i,i+r}^2
 x_{i-1,i+r+1}^2}{x_{i,i+r+1}^2 x_{i-1,i+r}^2} \Bigr)\ , \\ \nn
L_{n,i} &= - \frac{1}{2} \ln
 \Bigl(\frac{x_{i,i+m}^2}{x_{i,i+m+1}^2}\Bigr) \ln
\Bigl(\frac{x_{i+1,i+m+1}^2}{x_{i+m,i+2m}^2}\Bigr)\ .
\end{align}
For $n$ even, $n=2m$ they are
\begin{align}  \label{di2}
D_{n,i} &= - \sum_{r=2}^{m-2} {\rm Li}_2 \Bigl(1 - \frac{x_{i,i+r}^2
 x_{i-1,i+r+1}^2}{x_{i,i+r+1}^2 x_{i-1,i+r}^2} \Bigr) -
 \frac{1}{2}{\rm Li}_2\Bigl(1 - \frac{x_{i,i+m-1}^2
 x_{i-1,i+m}^2}{x_{i,i+m}^2 x_{i-1,i+m-1}^2} \Bigr)\ ,\\  \nn
L_{n,i} &= \frac{1}{4} \ln^2 \Big(\frac{x_{i,i+m}^2}{x_{i+1,i+m+1}^2}\Bigr)\ .
\end{align}
We have already seen that at four points and five points the general solution to
the Ward identity coincides with \re{ansatz1}. We now show that the ansatz
\re{ansatz1} is a solution of the Ward identity for arbitrary $n$.

First we observe that the dilogarithmic contributions in \re{di1} and \re{di2}
are functions of conformal cross-ratios of the form (\ref{crr}). They are
therefore invariant under conformal transformations and we have immediately
\be
K^\m D_{n,i} = 0\ .
\ee
For the logarithmic contributions we use the identity (\ref{Klog}). When $n$ is
odd we then find
\ba
&& \hspace*{-6mm} K^\m g_{n,i} = - 2\sum_{r=2}^{m-1}\bigl[ x_{i+r,i+r+1}^\m(\ln
 x_{i+1,i+r+1}^2 - \ln x_{i,i+r+1}^2) - x_{i,i+1}^\m(\ln x_{i,i+r}^2
 - \ln x_{i,i+r,2}^2)\bigr]
\\ \nn
&& -  x_{i+m,i+m+1}^\m(\ln x_{i+1,i+m+1}^2 - \ln
 x_{i+m,i+2m}^2)
- (x_{i+1,i+2m}^\m -
 x_{i+m,i+m+1}^\m)(\ln x_{i,i+m}^2 - \ln x_{i+m+1,i}^2)\, .
\ea
Changing variables term by term in the sum over $i$ one finds that only the $\ln
x_{i,i+2}^2$ terms remain and indeed (\ref{cwi}) is satisfied. The proof for $n$
even goes exactly the same way except that one obtains
\be
K^\m F_n = \sum_{i=1}^n \bigl[\ln x_{i,i+2}^2 (x_i^\m +
 x_{i+2}^\m - 2 x_{i+1}^\m) + \frac{1}{2} \ln x_{i,i+m}^2
 (x_{i+m-1,i+m+1}^\m - x_{i-1,i+1}^\m) \bigr]
\ee
and one has to use the fact that $n=2m$ to see that the $\ln x_{i,i+m}^2$ term
vanishes under the sum.

Thus we have seen that the BDS ansatz for the MHV amplitudes satisfies the
proposed conformal Ward identity for the Wilson loop. It would be interesting to
find out if the precise form of the conformally invariant dilogarithmic
contributions can be fixed by some other general properties of the Wilson loop.
The collinear behavior of the MHV amplitude discussed in Ref. \cite{Bern:1994zx}
indicates that such a guiding principle could come from the reduction from $n+1$
points to $n$ points. The requirement that the Wilson loop has a certain analytic
behavior, together with some mild assumption about the class of functions
involved, might be sufficient to fix this form uniquely.

\section{Conclusions}

In this paper we have demonstrated by explicit calculation that the duality
relation between light-like Wilson loop and four-gluon planar scattering
amplitude holds true in $\mathcal{N}=4$ SYM at weak coupling beyond one loop.
Making use of the nonabelian exponentiation theorem, we have shown that the
finite two-loop corrections to the Wilson loop dual to four-gluon amplitude
exponentiate with the prefactor given by the universal cusp anomalous dimension.
At first glance, this may be surprising since the same anomalous dimension
controls ultraviolet divergences that appear in the Wilson loop due to the
presence of the cusps on the integration contour. We have argued that this
property can be understood with the help of conformal symmetry. The cusp
singularities alter transformation properties of the Wilson loop under the
conformal $SO(2,4)$ transformations and produce anomalous contribution to the
conformal Ward identities. We proposed the all-loop form of the anomalous
conformal Ward identities and demonstrated that they uniquely fix the form of the
finite part (up to an additive constant) of the Wilson loop dual to four- and
five-gluon amplitudes, in complete agreement with the BDS conjecture for the
multi-gluon MHV amplitudes. Starting from six-gluon amplitudes, the conformal
symmetry is not sufficient to fix the finite part of the dual Wilson loops and
the conformal Ward identities should be supplemented by additional constraints on
analytical properties of Wilson loops. One may speculate on their possible
relation with the remarkable integrability symmetry of $\mathcal{N}=4$ SYM.

The conjectured planar gluon amplitudes/Wilson loop duality  offers a very clear
context in which to understand the appearance of conformal integrals in the
on-shell four-point MHV amplitude. These integrals were shown to be conformal up
to three loops in \cite{Drummond:2006rz} by writing the momenta in terms of dual
coordinates. It was found that the conformal pattern continues at four loops in
\cite{Bern:2006ew} and at five loops \cite{Bern:2007ct} was used as a guiding
principle in order to conjecture an amplitude whose consistency was checked with
various unitarity cuts. In \cite{DKS07} it was shown that certain integrals whose
coefficients vanish in the final expressions for the four-loop and conjectured
five-loop amplitudes are precisely those which do not have well-defined conformal
properties in four dimensions. Thus there is considerable evidence for a
conformal structure behind the perturbative four-point gluon amplitude. In
\cite{DKS07} it was argued that if the {\sl off-shell} regulated amplitude also
possessed such a conformal structure then basic properties of factorisation and
exponentiation of infrared divergences would fix the functional form of the
finite part (again up to an additive constant).

Since the conformal structure is best revealed in terms of the dual coordinate
variables $x_i$, it was referred to in \cite{DKS07} as `dual conformal
invariance'. In the on-shell four-point MHV amplitude this dual conformal
invariance is broken by the presence of a dimensional regulator. Similarly, for
the Wilson loop the ordinary conformal invariance of ${\cal N}=4$ SYM is broken
by the presence of a dimensional regulator. Thus the broken dual conformal
structure of the on-shell amplitude is mapped directly to the broken conformal
structure of the Wilson loop. The approach outlined here thus suggests the
possibility of directly exploiting the dual conformal structure of the on-shell
amplitude to understand the origin of its functional form. In this case one would
expect that also the form of the five-point amplitude would be fixed by dual
conformal symmetry. It would be very interesting to try to understand this in
terms of the five point integrals appearing there.

\section*{Acknowledgements}

We would like to thank A.~Belitsky, J.~Maldacena and D.~M\"uller for interesting
discussions. J.H.\ acknowledges the warm hospitality extended to him by the
Theory Group of the Dipartimento di Fisica, Universit\`{a} di Roma ``Tor
Vergata''. This research was supported in part by the French Agence Nationale de
la Recherche under grant ANR-06-BLAN-0142.

\appendix

\setcounter{section}{0} \setcounter{equation}{0}
\renewcommand{\theequation}{\Alph{section}.\arabic{equation}}

\section{Appendix}

For completeness, we give the explicit expressions for the integrals
\re{M-integrals} that appeared during the two-loop calculation
\begin{eqnarray}
M_{2}
&=& -\frac{\pi^2}{2} \ln(\alpha \bar{\alpha}) + 2 \rm{Li}_{3}(1)-
\rm{Li}_{3}\left(-\frac{\alpha}{\bar{\alpha}}\right)
-\rm{Li}_{3}\left(-\frac{\bar{\alpha}}{\alpha}\right) -
\ln\left(\frac{\alpha}{\bar{\alpha}}\right)
\left[ \rm{Li}_{2}(\alpha)-\rm{Li}_{2}(\bar{\alpha}) \right] \nonumber \\
M_{3}
&=&-\frac{49}{180}\pi^4 -\frac{1}{3}\pi^2 \left[ \ln^2(\alpha) +6 \ln(\alpha)
\ln(\bar{\alpha}) + \ln^2(\bar{\alpha}) \right] \nonumber
\\
&&-\frac{1}{12} \left[ \ln^4(\alpha) +\ln^4(\bar{\alpha}) +4
\ln(\bar{\alpha})\ln^3(\alpha)-18\ln^2(\alpha)\ln^2(\bar{\alpha})
 +4 \ln({\alpha})\ln^3(\bar{\alpha}) \right]
 \nonumber \\
 &&-4 \ln\left(\frac{\alpha}{\bar{\alpha}}\right) \left[
\rm{Li}_{3}\left(\alpha\right) - \rm{Li}_{3}\left(\bar{\alpha}\right) \right] + 8
\left[ \rm{Li}_{4}\left(\alpha\right)+ \rm{Li}_{4}\left(\bar{ \alpha}\right)
\right] \nonumber
\end{eqnarray}
where $\bar\alpha=1-\alpha$ and ${\rm Li}_{n}(z)$ (with $n=2,3,4$) are
polylogarithms \cite{Lewin}.


\begin{thebibliography}{99}

\bibitem{bds05} Z.~Bern, L.~J.~Dixon and V.~A.~Smirnov,
  ``Iteration of planar amplitudes in maximally supersymmetric Yang-Mills
  theory at three loops and beyond,''
  Phys.\ Rev.\  D {\bf 72} (2005) 085001
  [arXiv:hep-th/0505205].

\bibitem{P80}
  A.~M.~Polyakov,
  ``Gauge Fields As Rings Of Glue,''
  Nucl.\ Phys.\  B {\bf 164} (1980) 171.

\bibitem{KR87}
  G.~P.~Korchemsky and A.~V.~Radyushkin,
  ``Loop Space Formalism And Renormalization Group For The Infrared Asymptotics
  Of QCD,''
  Phys.\ Lett.\  B {\bf 171} (1986) 459;
  ``Renormalization of the Wilson Loops Beyond the Leading Order,''
  Nucl.\ Phys.\  B {\bf 283} (1987) 342.

\bibitem{5point}
  F.~Cachazo, M.~Spradlin and A.~Volovich,
  ``Iterative structure within the five-particle two-loop amplitude,''
  Phys.\ Rev.\  D {\bf 74} (2006) 045020
  [arXiv:hep-th/0602228];
\\ Z.~Bern, M.~Czakon, D.~A.~Kosower, R.~Roiban and V.~A.~Smirnov,
  ``Two-loop iteration of five-point N = 4 super-Yang-Mills amplitudes,''
  Phys.\ Rev.\ Lett.\  {\bf 97} (2006) 181601
  [arXiv:hep-th/0604074].

\bibitem{am07}L.~F.~Alday and J.~Maldacena,
  ``Gluon scattering amplitudes at strong coupling,'' JHEP {\bf 0706} (2007) 064
  [arXiv:0705.0303 [hep-th]].

\bibitem{AdS} J~.M.~Maldacena, ``The large N limit of superconformal field theories
and supergravity,''
Adv. Theor. Math. Phys. {\bf 2} (1998) 231;\\
S.~S.~Gubser, I.~R.~Klebanov and A.~M.~Polyakov, ``Gauge theory correlators from
non-critical string theory,''
Phys. Lett. B {\bf 428} (1998) 105;\\
E.~Witten, ``Anti-de Sitter space and holography,'' Adv. Theor. Math. Phys. {\bf
2} (1998) 253.

\bibitem{Kallosh:1998ji}
  R.~Kallosh and A.~A.~Tseytlin,
  ``Simplifying superstring action on AdS(5) x S(5),''
  JHEP {\bf 9810} (1998) 016
  [arXiv:hep-th/9808088].

\bibitem{M98}
  J.~M.~Maldacena,
  ``Wilson loops in large N field theories,''
  Phys.\ Rev.\ Lett.\  {\bf 80} (1998) 4859
  [arXiv:hep-th/9803002];
  \\
  S.~J.~Rey and J.~T.~Yee,
  ``Macroscopic strings as heavy quarks in large N gauge theory and  anti-de
  Sitter supergravity,''
  Eur.\ Phys.\ J.\  C {\bf 22} (2001) 379
  [arXiv:hep-th/9803001].

\bibitem{Kr02}
  M.~Kruczenski,
  ``A note on twist two operators in N = 4 SYM and Wilson loops in Minkowski
  signature,''
  JHEP {\bf 0212} (2002) 024
  [arXiv:hep-th/0210115];
  \\
  Yu.~Makeenko,
  ``Light-cone Wilson loops and the string/gauge correspondence,''
  JHEP {\bf 0301} (2003) 007
  [arXiv:hep-th/0210256].

\bibitem{GPK} S.~S.~Gubser, I.~R.~Klebanov and A.~M.~Polyakov,
  ``A semi-classical limit of the gauge/string correspondence,''
  Nucl.\ Phys.\  B {\bf 636} (2002) 99
  [arXiv:hep-th/0204051].

\bibitem{KRTT07}
  M.~Kruczenski, R.~Roiban, A.~Tirziu and A.~A.~Tseytlin,
  ``Strong-coupling expansion of cusp anomaly and gluon amplitudes from quantum
  open strings in AdS${}_5$ x S${}^5$,''
  arXiv:0707.4254 [hep-th].

\bibitem{DKS07}
J.~M.~Drummond, G.~P.~Korchemsky and E.~Sokatchev, ``Conformal properties of
four-gluon planar amplitudes and Wilson loops,'' arXiv:0707.0243 [hep-th].

\bibitem{BHT07}
  A.~Brandhuber, P.~Heslop and G.~Travaglini,
  ``MHV Amplitudes in N=4 Super Yang-Mills and Wilson Loops,''
  arXiv:0707.1153 [hep-th].

\bibitem{Bern:1997nh}
  Z.~Bern, J.~S.~Rozowsky and B.~Yan,
  ``Two-loop four-gluon amplitudes in N = 4 super-Yang-Mills,''
  Phys.\ Lett.\  B {\bf 401} (1997) 273
  [arXiv:hep-ph/9702424];\\
  C.~Anastasiou, Z.~Bern, L.~J.~Dixon and D.~A.~Kosower,
  ``Planar amplitudes in maximally supersymmetric Yang-Mills theory,''
  Phys.\ Rev.\ Lett.\  {\bf 91} (2003) 251602
  [arXiv:hep-th/0309040].

\bibitem{Gatheral}
  G. Sterman, in AIP Conference Proceedings Tallahassee, Perturbative Quantum
   Chromodynamics, eds. D. W. Duke, J. F. Owens, New York, 1981, p. 22; \\
   J.~G.~M.~Gatheral,
  ``Exponentiation Of Eikonal Cross-Sections In Nonabelian Gauge Theories,''
  Phys.\ Lett.\  B {\bf 133} (1983) 90;\\
  J.~Frenkel and J.~C.~Taylor,
  ``Nonabelian Eikonal Exponentiation,''
  Nucl.\ Phys.\  B {\bf 246} (1984) 231.

\bibitem{K88}
G.~P.~Korchemsky, ``Asymptotics of the Altarelli-Parisi-Lipatov Evolution Kernels
of Parton Distributions,'' Mod.\ Phys.\ Lett.\  A {\bf 4} (1989) 1257; \\
G.~P.~Korchemsky and G.~Marchesini, ``Structure function for large x and
renormalization of Wilson loop,''
  Nucl.\ Phys.\  B {\bf 406} (1993) 225
  [arXiv:hep-ph/9210281].

\bibitem{KK92}
  I.~A.~Korchemskaya and G.~P.~Korchemsky,
  ``On light-like Wilson loops,''
  Phys.\ Lett.\  B {\bf 287} (1992) 169;\\
  A.~Bassetto, I.~A.~Korchemskaya, G.~P.~Korchemsky and G.~Nardelli,
  ``Gauge invariance and anomalous dimensions of a light cone Wilson loop in
  lightlike axial gauge,''
  Nucl.\ Phys.\  B {\bf 408} (1993) 62
  [arXiv:hep-ph/9303314].

\bibitem{BGK03}
  A.~V.~Belitsky, A.~S.~Gorsky and G.~P.~Korchemsky,
  ``Gauge/string duality for QCD conformal operators,''
  Nucl.\ Phys.\  B {\bf 667} (2003) 3
  [arXiv:hep-th/0304028].

\bibitem{KLOV} A.~V.~Kotikov, L.~N.~Lipatov, A.~I.~Onishchenko and V.~N.~Velizhanin,
  ``Three-loop universal anomalous dimension of the Wilson operators in N =  4
  SUSY Yang-Mills model,''
  Phys.\ Lett.\  B {\bf 595} (2004) 521
  [Erratum-ibid.\  B {\bf 632} (2006) 754]
  [arXiv:hep-th/0404092].

\bibitem{Braun:2003rp}
  V.~M.~Braun, G.~P.~Korchemsky and D.~Mueller,
  ``The uses of conformal symmetry in QCD,''
  Prog.\ Part.\ Nucl.\ Phys.\  {\bf 51} (2003) 311
  [arXiv:hep-ph/0306057].

\bibitem{Mueller:1993hg}
  D.~M\"uller,
  ``Conformal constraints and the evolution of the nonsinglet meson
  distribution amplitude,''
  Phys.\ Rev.\  D {\bf 49} (1994) 2525;\\
 D.~M\"uller,
  ``Restricted conformal invariance in QCD and its predictive power for
  virtual two-photon processes,''
  Phys.\ Rev.\  D {\bf 58} (1998) 054005
  [arXiv:hep-ph/9704406]; \\
  A.~V.~Belitsky and D.~M\"uller,
  ``Broken conformal invariance and spectrum of anomalous dimensions in
  {QCD},''
  Nucl.\ Phys.\  B {\bf 537} (1999) 397
  [arXiv:hep-ph/9804379].

\bibitem{Bern:1994zx}
  Z.~Bern, L.~J.~Dixon, D.~C.~Dunbar and D.~A.~Kosower,
  ``One loop n point gauge theory amplitudes, unitarity and collinear limits,''
  Nucl.\ Phys.\  B {\bf 425}, 217 (1994)
  [arXiv:hep-ph/9403226].

\bibitem{Drummond:2006rz}
  J.~M.~Drummond, J.~Henn, V.~A.~Smirnov and E.~Sokatchev,
  ``Magic identities for conformal four-point integrals,''
  JHEP {\bf 0701} (2007) 064
  [arXiv:hep-th/0607160].

\bibitem{Bern:2006ew}
  Z.~Bern, M.~Czakon, L.~J.~Dixon, D.~A.~Kosower and V.~A.~Smirnov,
  ``The Four-Loop Planar Amplitude and Cusp Anomalous Dimension in Maximally
  Supersymmetric Yang-Mills Theory,''
  Phys.\ Rev.\  D {\bf 75}, 085010 (2007)
  [arXiv:hep-th/0610248].

\bibitem{Bern:2007ct}
  Z.~Bern, J.~J.~M.~Carrasco, H.~Johansson and D.~A.~Kosower,
  ``Maximally supersymmetric planar Yang-Mills amplitudes at five loops,''
  arXiv:0705.1864 [hep-th].

\bibitem{Lewin}
L.~Lewin, ``Polylogarithms and Associated Functions,'' North-Holland-New York,
1981.


\end{thebibliography}
\end{document}